\documentclass[conference]{IEEEtran}
\IEEEoverridecommandlockouts
\usepackage{cite}
\usepackage{amsmath,amssymb,amsfonts}
\usepackage{algorithmic}
\usepackage{graphicx}
\usepackage{textcomp}
\usepackage{xcolor}
\def\BibTeX{{\rm B\kern-.05em{\sc i\kern-.025em b}\kern-.08em
    T\kern-.1667em\lower.7ex\hbox{E}\kern-.125emX}}
\begin{document}

\title{kNN Classification of Malware Data Dependency Graph Features}

\author{\IEEEauthorblockN{John Musgrave}
\IEEEauthorblockA{\textit{Computer Science Department} \\
\textit{The College of Wooster}\\
Wooster, OH \\
jmusgrave@wooster.edu}
\and
\IEEEauthorblockN{Anca Ralescu}
\IEEEauthorblockA{\textit{Department of Computer Science} \\
\textit{University of Cincinnati}\\
Cincinnati, OH \\
ralescal@ucmail.uc.edu}
}

\maketitle

\begin{abstract}
Explainability in classification results are dependent upon the features used for classification.  Data dependency graph features representing data movement are directly correlated with operational semantics, and subject to fine grained analysis.  This study obtains accurate classification from the use of features tied to structure and semantics.  By training an accurate model using labeled data, this feature representation of semantics is shown to be correlated with ground truth labels.  This was performed using non-parametric learning with a novel feature representation on a large scale dataset, the Kaggle 2015 Malware dataset.  The features used enable fine grained analysis, increase in resolution, and explainable inferences.  This allows for the body of the term frequency distribution to be further analyzed and to provide an increase in feature resolution over term frequency features.  This method obtains high accuracy from analysis of a single instruction, a method that can be repeated for additional instructions to obtain further increases in accuracy.  This study evaluates the hypothesis that the semantic representation and analysis of structure are able to make accurate predications and are also correlated to ground truth labels. Additionally, similarity in the metric space can be calculated directly without prior training.  Our results provide evidence that data dependency graphs accurately capture both semantic and structural information for increased explainability in classification results.
\end{abstract}


\section{Introduction}
In recent years many studies have focused on the applications of machine learning to malware analysis.  However, increases in interpretability are limited by the features used in various classifiers.  This study explores the use of features constructed from graphs of data dependencies \cite{souri2018state}.

Many graph features for malware analysis are focused on control flow graph features.  When data flow tracking analysis is performed, it is often evaluated in a top-down approach and compares data flow and data dependencies between functions in high level languages \cite{bruschi2006detecting}.

Explainable approaches to operational semantics remain an open question.
It remains unclear why data movement instructions have few features that allow for increases in resolution, given their prevalence of frequency.  Recent advances in large language models do not directly address explainability \cite{xu2022systematic}.

Data dependency graphs as features represent both operational semantics and structural properties of binaries, can be constructed in a bottom-up approach, and offer a potential increase in explainability.  The purpose of this study is to measure the accuracy of classifiers trained on these feature representations.

\begin{table}[h]
    \centering
    \begin{tabular}{|c|c|}
    \hline
         Term (x86/64 Opcode) & \# Of Occurrences \\
         \hline
        MOV & 61,312,709 \\
        PUSH & 26,067,804 \\
        CALL & 12,158,797 \\
        IMUL & 8,174,383 \\
        POP & 8,157,787 \\
        MUL & 8,114,871 \\
        NOP & 7,307,405 \\
        XOR & 7,171,442 \\
        CMP & 6,997,102 \\
        ADD & 6,404,277 \\
        LEA & 6,252,131 \\
        STD & 5,899,174 \\
        JZ & 5,124,385 \\
        TEST & 4,682,198 \\
        JNZ & 3,555,014 \\
        SUB & 3,461,608 \\
        CLD & 3,051,495 \\
        AND & 2,387,916 \\
        INC & 1,708,580 \\
        OR & 1,635,851 \\
        INT & 1,288,737 \\
        MOVZX & 1,227,559 \\
        XCHG & 1,122,534 \\
        DEC & 931,127 \\
        SHL & 827,192 \\
        \hline
    \end{tabular}
    \caption{Top 25 opcodes in the Kaggle 2015 dataset ordered by frequency in descending order}
    \label{tab:opcode_frequencies}
\end{table}

\subsection{Related Work}
The Microsoft Kaggle Malware dataset has been used successfully for many studies on malware analysis and classification \cite{kebede2017classification}, \cite{ronen2018microsoft}.

Several studies on malware have focused on control flow graph representations of programs and their use in classification. A number of studies explored the use of static features of file metadata.  Decision trees for the classification of Windows PE files have been effective for classification.  Subsequent studies have used ensemble methods, random forests, and support vector machines, with features extracted from file headers in Trojan malware 
\cite{bruschi2006detecting}, \cite{cesare2010fast}, \cite{cesare2013control}, \cite{cesare2010classification}, \cite{shafiq2009pe}, \cite{siddiqui2008detecting}, \cite{witten1999weka}, 
\cite{chandrasekaran2021malware}.

\begin{figure*}[h]
    \centering
    \includegraphics[width=0.95\textwidth]{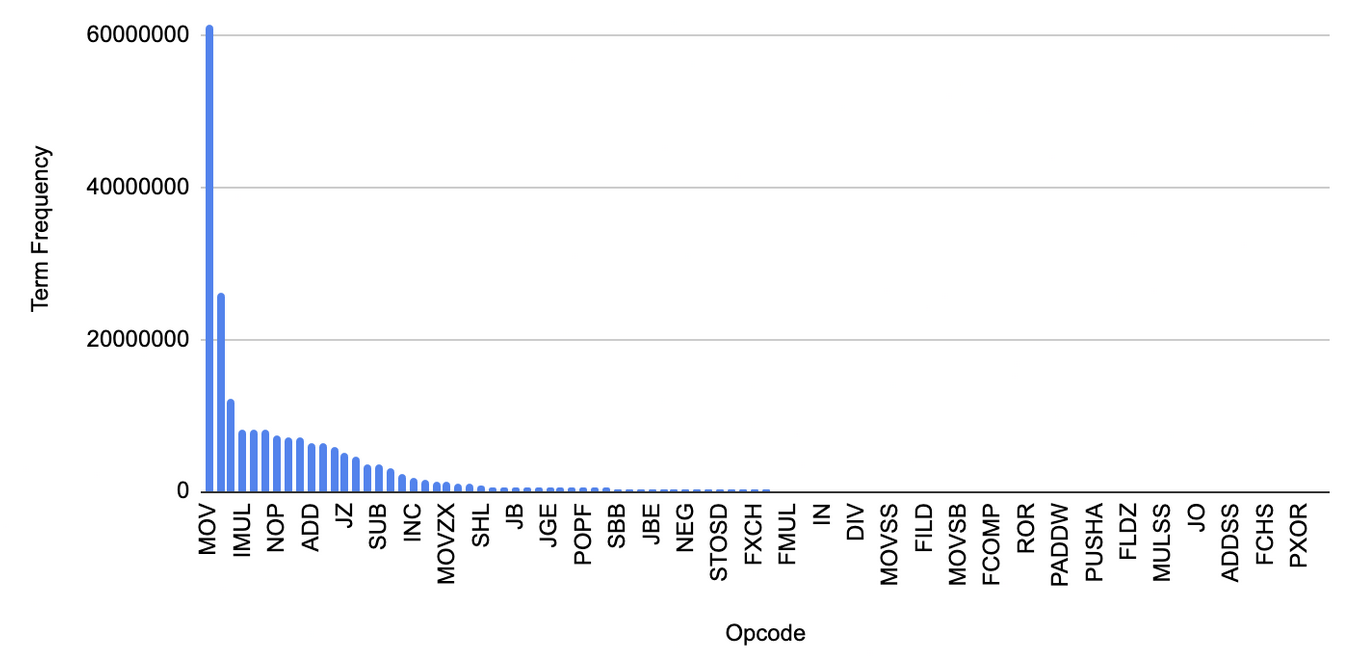}
    \caption{Histogram of opcode frequencies in the Microsoft 2015 Kaggle malware dataset}
    \label{fig:opcode_histogram}
\end{figure*}

Several studies have focused on function abstraction semantics through decompilation.  LeDoux et al. represented a program as a graph of function abstractions obtained from reverse engineering and used semantic hashing as a measurement of similarity.  However, this study did not take a bottom up approach, and basic block features were specifically not considered.  There may be many equivalent programs for a given malware binary, and whether semantic function abstractions in a high level language are correlated to lower level binary representations is an open question.  In a similar manner, Alrabaee et al. have used a $tf-idf$ representation with Hidden Markov Models and graph kernels to obtain a graph of semantic function abstractions for a program.  This was accomplished by constructing a Bayesian network for each of the features collected  \cite{ledoux2013functracker}, \cite{alrabaee2018fossil}.

In a previous study we have performed a comparison of various classifiers and clustering algorithms on a dataset of $tf-idf$ features. In a previous study, we have been able to measure behavioral overlap between individual program samples. We intend to show increases in explainability through the use of a novel feature representation.  Additionally we perform this analysis on a larger dataset of malware, the Microsoft Kaggle malware dataset.  \cite{musgrave2020semantic}, \cite{musgrave2024search}.

\subsection{Outline}
Section 2 covers the details of the experiment performed, including method of data collection and pre-processing.  Section 3 covers experimental results.  Section 4 contains an analysis of the results, and a discussion of the findings and their implications.  Section 5 contains a summary and conclusion.

\section{Experiment}
This section outlines the methods used in performing our experiments.

\subsection{Data Pre-Processing}
The dataset used in this study was composed by processing the Microsoft 2015 Kaggle Malware dataset. This dataset uses the output of the Interactive Disassembler or IDA Pro, and contains the text sections of over 10,000 binary samples in the training set, along with their associated class labels. Each sample contains the disassembled binary, and the output contains both the hex values and x86 assembly instructions for the decompiled binary. Each sample has a corresponding hash value as an identifier. Each sample in the dataset belongs to one of 9 classes of malware, and there are no benign samples present in the
dataset \cite{ronen2018microsoft}.

As a motivating example, we present the results of a term frequency measurement on the dataset. Using the $x86/64$ instruction set we have constructed a term dictionary that includes each instruction opcode as a term. The frequency of each opcode's occurrence was measured across the dataset, and this is shown in Table \ref{tab:opcode_frequencies} and Figure \ref{fig:opcode_histogram}. The overwhelming presence of data movement instructions as seen in Table \ref{tab:opcode_frequencies} are the primary motivation for additional pre-processing and feature extraction.  $tf-idf$ features often do not take into consideration the semantics of data movement.

While a majority of the feature pre-processing and analysis was performed in Python directly, additional methods in the $scikit-learn$ Python library were used for the implementation of various algorithms \cite{pedregosa2011scikit}.

Our dataset was segmented into basic blocks for further analysis, as outlined in previous work. This is necessary for additional adjustments in feature granularity. Segmentation was performed on instructions which require control transfer, such as any number of jump instructions or the $call$ instruction. Instructions using the stack such as $push$ were not evaluated, and control flow analysis was not the primary focus of this study. A single segment in the dataset corresponds to a contiguous segment of instructions, which represents an atomic unit of operational semantics. Instructions re-ordered by the processor must maintain semantic coherence, and this is done by maintaining the coherence of dependencies. Therefore, a program can be represented as a collection of these segments, which allows for further increases in resolution \cite{musgrave2022latent}.

Data dependency graphs were extracted for each segment and hashed for isomorphism to construct a set of unique hashes as outlined in previous work. Each program segment corresponds to a single data dependency graph, and a single graph of data dependencies has an associated program segment. Data operands are added as nodes in the graph, and a dependency is added as a graph edge.  Frequency of graph hashes is not considered in our representation, and each set is filtered for a unique set of isomorphic hashes. While data dependencies exist between many arithmetic instructions in the term dictionary, our focus was on data movement instructions, since this term had the highest frequency.  Expanding our work to operands of other terms is an area to be developed in future studies, as it may provide increases in resolution. Similarly the differentiation of indirection may provide increases in accuracy and resolution, we have not included an analysis of data indirection in the features constructed, which are evaluated as separate
operands \cite{musgrave2024empirical}, \cite{musgrave2022novel}.

Once the dataset has been segmented and data dependency graphs are extracted each graph is hashed for uniqueness using the Weisfeiler-Lehman graph hashing algorithm.  This algorithm ensures that two graphs which are isomorphic produce an identical hash value.  Each unique hash is then recorded once in a list resulting in features we have labeled a DDG Fingerprint \cite{shervashidze2011weisfeiler}, \cite{musgrave2022novel}, \cite{musgrave2024search}.

A single program in the dataset corresponds to a set of hashes, each set containing a unique list of categorical values.  The categorical value corresponds to the pattern of data movement ensured by the hashing algorithm for graph isomorphism.  Our dataset contains 10,617 sets of hashes, and there are 74,872 isomorphically unique patterns across the entire dataset.

\subsection{Hamming Space}
We construct a metric space using the Hamming Distance between the vectors of categorical values. This is done by creating a one-hot encoding for each hash value. The resulting metric space is a high dimensional space where the distance to each vector can be calculated based on the Hamming Distance. This approach is outlined in prior work \cite{musgrave2024search}.

The primary motivation for the use of Hamming Distance is that it provides a distance for categorical features. Our study analyzes the formation of clusters in the space constructed using this distance metric.

\subsection{k-Nearest Neighbors Classifier}
We train a classifier using the k-Nearest Neighbors (kNN) algorithm. The kNN algorithm classifies an example based on the majority class of its k-nearest neighbors in the feature space. For our study, we used the training labels provided in the Kaggle 2015 malware dataset to compose a labeled dataset. This dataset was split into training and test sets to be used for classification, which are 75\% training and 25\% for the test set. The training set was composed of 7,962 samples, and the test set was composed of 2,655 samples.  The classifier measured were trained on the labeled training set, which has 9 class labels of malicious programs.  The test set did not use class labels, and the computed pairwise distances were maintained from the complete dataset.  This is done using the $scikit-learn$ Python library.  The accuracy metrics presented were measured from predictions of the classifiers over the test set \cite{pedregosa2011scikit}.

\subsection{Accuracy Metrics}
We evaluate the results from the k-Nearest Neighbors classifier by selecting a set of accuracy metrics.  We evaluate each classifier in terms of its accuracy, precision, recall, specificity, and F1 score, defined by the following:

\[ Accuracy = \frac{TP+TN}{TP+TN+FP+FN} \]
\[ Precision = \frac{TP}{TP+FP} \]
\[ Recall = \frac{TP}{TP+FN} \]
\[Specificity = \frac{TN}{FP+TN}\]
\[ F1 = \frac{2*Precision*Recall}{Precision+Recall} = \frac{2*TP}{2*TP+FP+FN}
 \]

This is performed over a variety of values of $k$ for comparison purposes between kNN classifiers.
 
\begin{figure*}
    \centering
    \includegraphics[width=1\textwidth, height=0.65\textwidth]{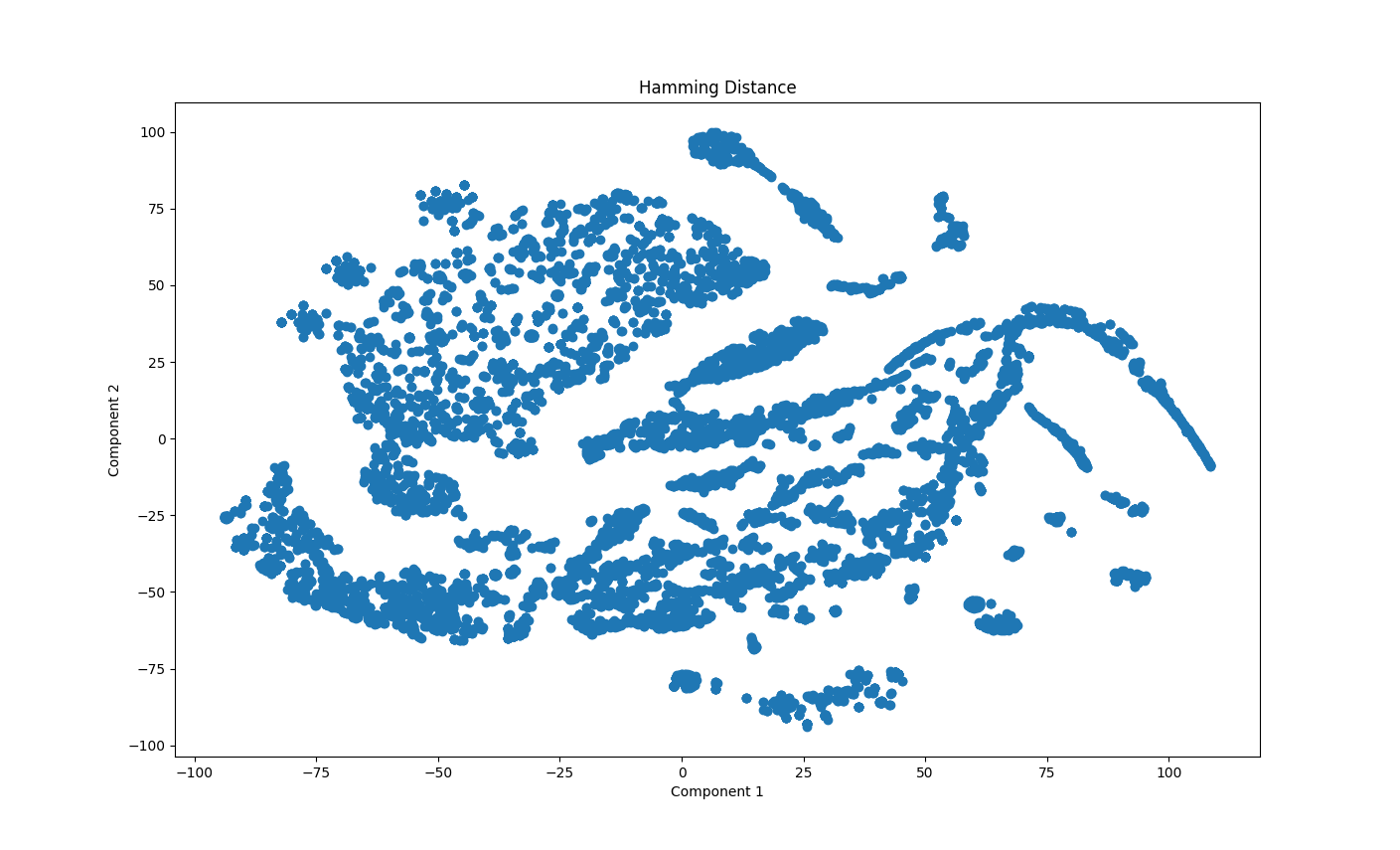}
    \caption{Kaggle Malware Dataset Hamming Space.  This figure shows the dataset projected from a high dimensional metric space with Hamming Distance to 2 dimensions using t-SNE.}
    \label{fig:enter-label}
    
    \centering
    \includegraphics[width=1\textwidth, height=0.65\textwidth]{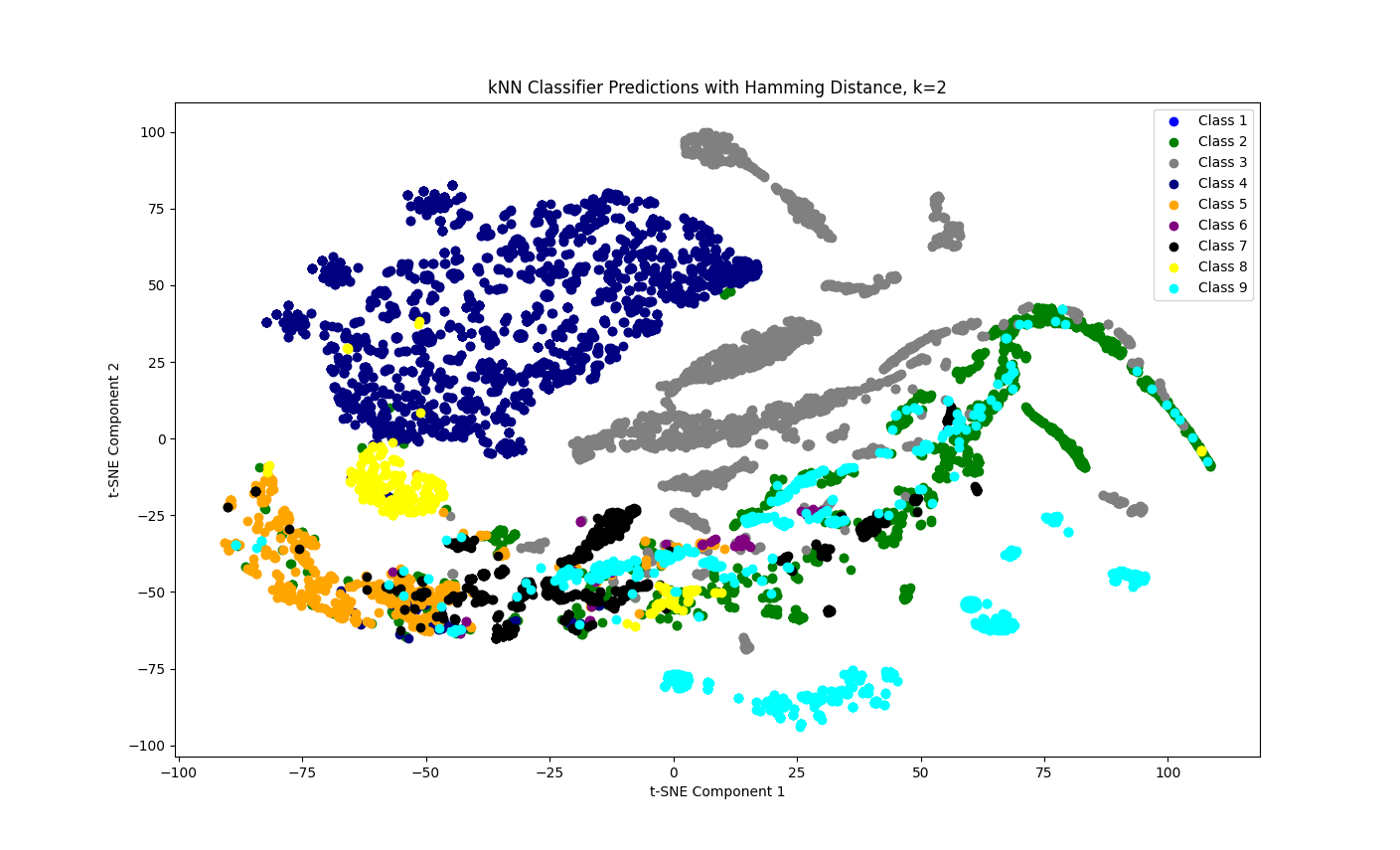}
    \caption{kNN Predictions.  Pictured above is the composition in 2 dimensions of the high dimensional hamming space with predicted class labels indicated by the color of each data point.}
    \label{fig:enter-label}
\end{figure*}
 \begin{figure*}
    \centering
    \includegraphics[width=1\textwidth, height=0.65\textwidth]{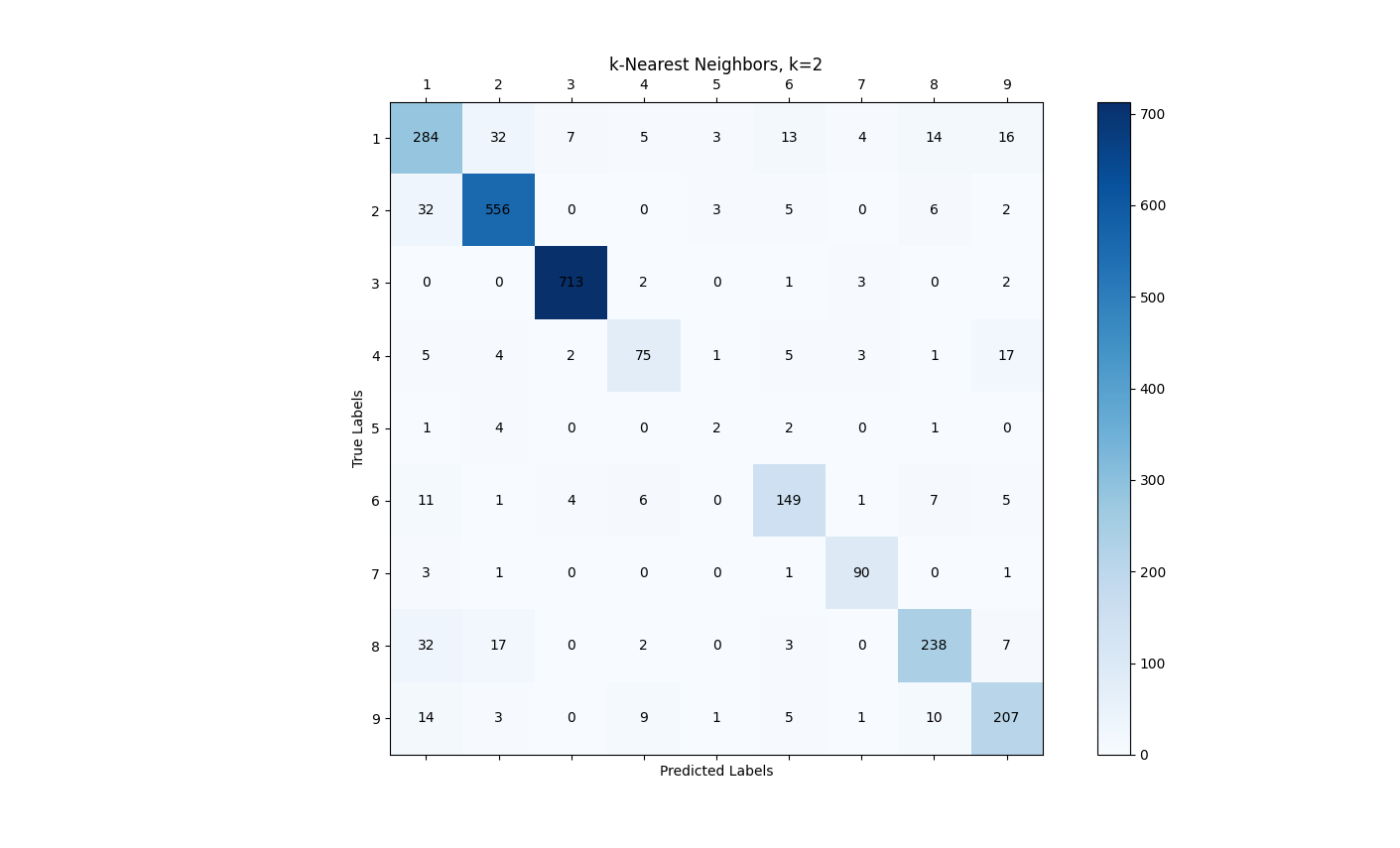}
    \caption{Confusion matrix for k-Nearest Neighbors classifier with k=2.}
    \vspace{20pt}
    \label{fig:enter-label}
    
    \begin{tabular}{|c|c|c|c|c|c|}
        \hline
        Class & Accuracy & Precision & Recall & Specificity & F1 Score \\
        \hline
        1 & 0.9345 & 0.6535 & 0.6935 & 0.9875 & 0.6730 \\
        2 & 0.9667 & 0.8275 & 0.8965 & 0.9856 & 0.8605 \\
        3 & 0.9898 & 0.9630 & 0.9930 & 0.9907 & 0.9780 \\
        4 & 0.9780 & 0.8935 & 0.8240 & 0.9960 & 0.8570 \\
        5 & 0.9876 & 0.6665 & 0.2800 & 0.9956 & 0.3905 \\
        6 & 0.9746 & 0.8070 & 0.8105 & 0.9887 & 0.8087 \\
        7 & 0.9890 & 0.9230 & 0.9375 & 0.9930 & 0.9302 \\
        8 & 0.9497 & 0.7440 & 0.7745 & 0.9820 & 0.7590 \\
        9 & 0.9690 & 0.8035 & 0.8205 & 0.9926 & 0.8119 \\
        \hline
    \end{tabular}
    \caption{Accuracy metrics per class for kNN classifier with k=2.  Precision, recall, specificity, and F1 scores are displayed for each class.}
\end{figure*}

\begin{figure*}
    \centering
    \includegraphics[width=1\textwidth, height=0.65\textwidth]{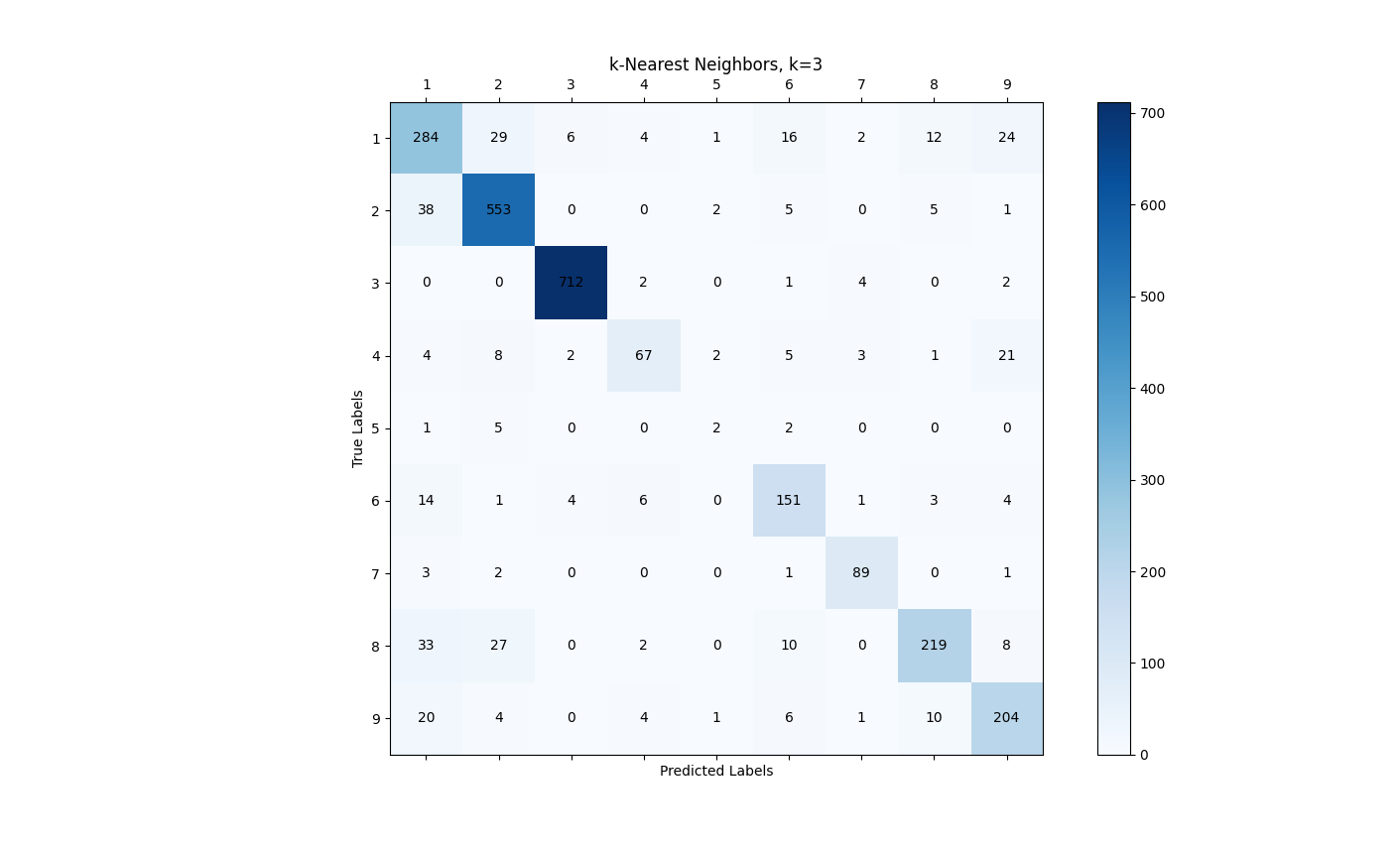}
    \caption{Confusion Matrix for kNN classifier, k=3}
    \label{fig:enter-label}

    \includegraphics[width=1\textwidth, height=0.65\textwidth]{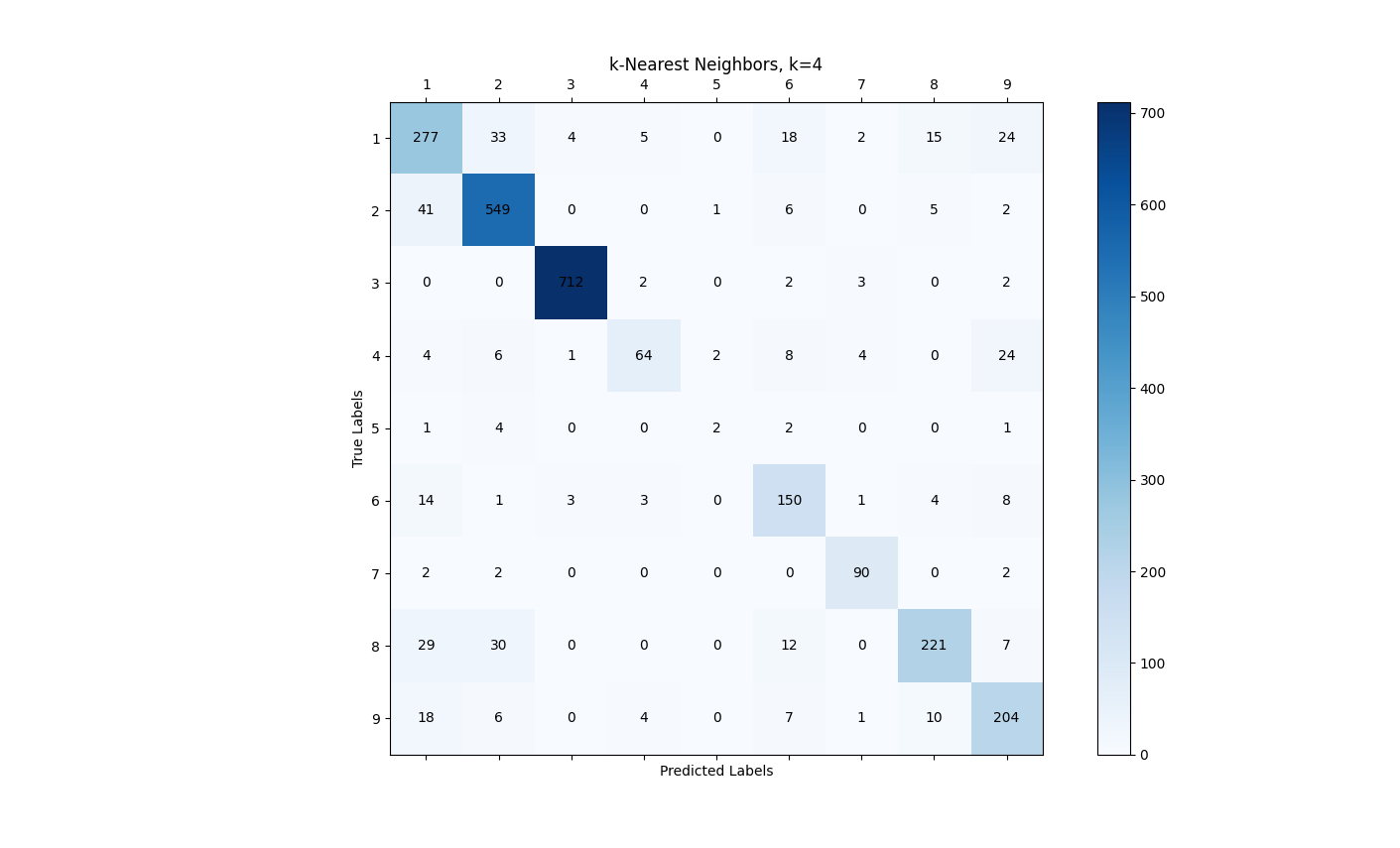}
    \caption{Confusion Matrix for kNN classifier, k=4}
    \label{fig:enter-label}
\end{figure*}

\begin{table*}[h]
    \centering
    \begin{tabular}{|c|c|c|c|c|c|}
        \hline
        Class & Accuracy & Precision & Recall & Specificity & F1 Score \\
        \hline
        1 & 0.9320 & 0.6553 & 0.6923 & 0.9871 & 0.6733 \\
        2 & 0.9650 & 0.8158 & 0.8957 & 0.9846 & 0.8541 \\
        3 & 0.9892 & 0.9616 & 0.9916 & 0.9903 & 0.9764 \\
        4 & 0.9764 & 0.8766 & 0.8148 & 0.9950 & 0.8447 \\
        5 & 0.9875 & 0.6471 & 0.2857 & 0.9956 & 0.3953 \\
        6 & 0.9738 & 0.7989 & 0.8053 & 0.9879 & 0.8021 \\
        7 & 0.9885 & 0.9194 & 0.9375 & 0.9926 & 0.9284 \\
        8 & 0.9495 & 0.7421 & 0.7733 & 0.9818 & 0.7574 \\
        9 & 0.9686 & 0.8023 & 0.8180 & 0.9924 & 0.8101 \\
        \hline
    \end{tabular}
    \caption{Accuracy metrics per class for kNN classifier with k=3.  Precision, recall, specificity, and F1 scores are displayed for each class.}
    \label{tab:metrics}
\end{table*}

\begin{table*}[h]
    \centering
    \begin{tabular}{|c|c|c|c|c|c|}
        \hline
        Class & Accuracy & Precision & Recall & Specificity & F1 Score \\
        \hline
        1 & 0.9326 & 0.6641 & 0.6892 & 0.9874 & 0.6761 \\
        2 & 0.9656 & 0.8225 & 0.8935 & 0.9851 & 0.8566 \\
        3 & 0.9893 & 0.9616 & 0.9916 & 0.9903 & 0.9764 \\
        4 & 0.9759 & 0.8571 & 0.7795 & 0.9945 & 0.8167 \\
        5 & 0.9876 & 0.6667 & 0.2857 & 0.9957 & 0.3976 \\
        6 & 0.9740 & 0.8021 & 0.7895 & 0.9881 & 0.7958 \\
        7 & 0.9887 & 0.9231 & 0.9375 & 0.9928 & 0.9302 \\
        8 & 0.9494 & 0.7407 & 0.7593 & 0.9818 & 0.7500 \\
        9 & 0.9688 & 0.8031 & 0.8160 & 0.9925 & 0.8095 \\
        \hline
    \end{tabular}
    \caption{Accuracy metrics per class for kNN classifier with k=4.  Precision, recall, specificity, and F1 scores are displayed for each class.}
    \label{tab:metrics}
\end{table*}

\section{Results}
Table 1 contains a partial list of the $x86/64$ instruction opcodes taken from the term dictionary. This table shows the frequency of occurrence for the most frequent instructions in the dataset. We can see that the '$mov$' instruction has the highest frequency.

Figure 1 shows the term frequency data plotted as a frequency histogram in order to see the distribution of term frequency. We can see that the '$mov$' instruction has the highest frequency in the Kaggle malware dataset, with over 61 Million occurrences.

Figure 2 shows a scatter plot of complete dataset in Hamming Space. Since the Hamming vectors are high dimensional, this figure shows a projection of the high dimensional space in 2 dimensions using t-SNE. The total dimensionality of the space is 74,872 \cite{hinton2002stochastic}.

Figure 3 shows a scatter plot of the dataset with class predictions from the kNN classifier for each data point. The Kaggle malware dataset is composed of 9 distinct malware classes with no benign samples.

Figure 4 shows a confusion matrix for the k-NN classifier with k=2. The accuracy metrics were used to measure the classifier's performance across the test set of n samples. True labels are presented on the vertical axis, and predicted labels on the horizontal axis. We return to an analysis of the selected accuracy metrics in the following section.

Figure 5 and the associated table presents accuracy metrics for the same classifier.  Total accuracy is the true positive plus false positive divided by the total, in order to exclude mispredictions. In order to calculate the total accuracy we can take the true positive value of 2,314 true predictions and divide it by the size of the test set, 2,657. More detail on the total accuracy of classifiers are presented in table 3.

Figure 6 and 7 present results for kNN classifiers with k values 3 and 4 in a confusion matrix. Both classifiers have a total accuracy of 85\% or better.  Tables 2 and 3 present accuracy metrics for kNN classifiers with k=3 and k=4, respectively. The accuracy for each class, along with the precision, recall, specificity, and F1 score are presented.

Figure 8 shows the decision boundary of the k-NN classifier in the Hamming Space.

Figure 9 shows the total accuracy of the classifier and the impact of k as the value of k increases. We can see that as k increases, the total accuracy of the classifier decreases. The total accuracy is calculated by the number of true predictions over the total number predictions when measured over the test set.

Additional figures are presented in the appendix section. Figure 9 shows a confusion matrix for k=5. Table 5 presents accuracy metrics for a kNN classifier with k=5. Figure 10 shows a scatter plot of the dataset with true class labels. Figure 11 shows a scatter plot of predictions from the kNN classifier with k=5.

\begin{figure*}
    \centering
    \includegraphics[width=1\textwidth, height=0.6\textwidth]{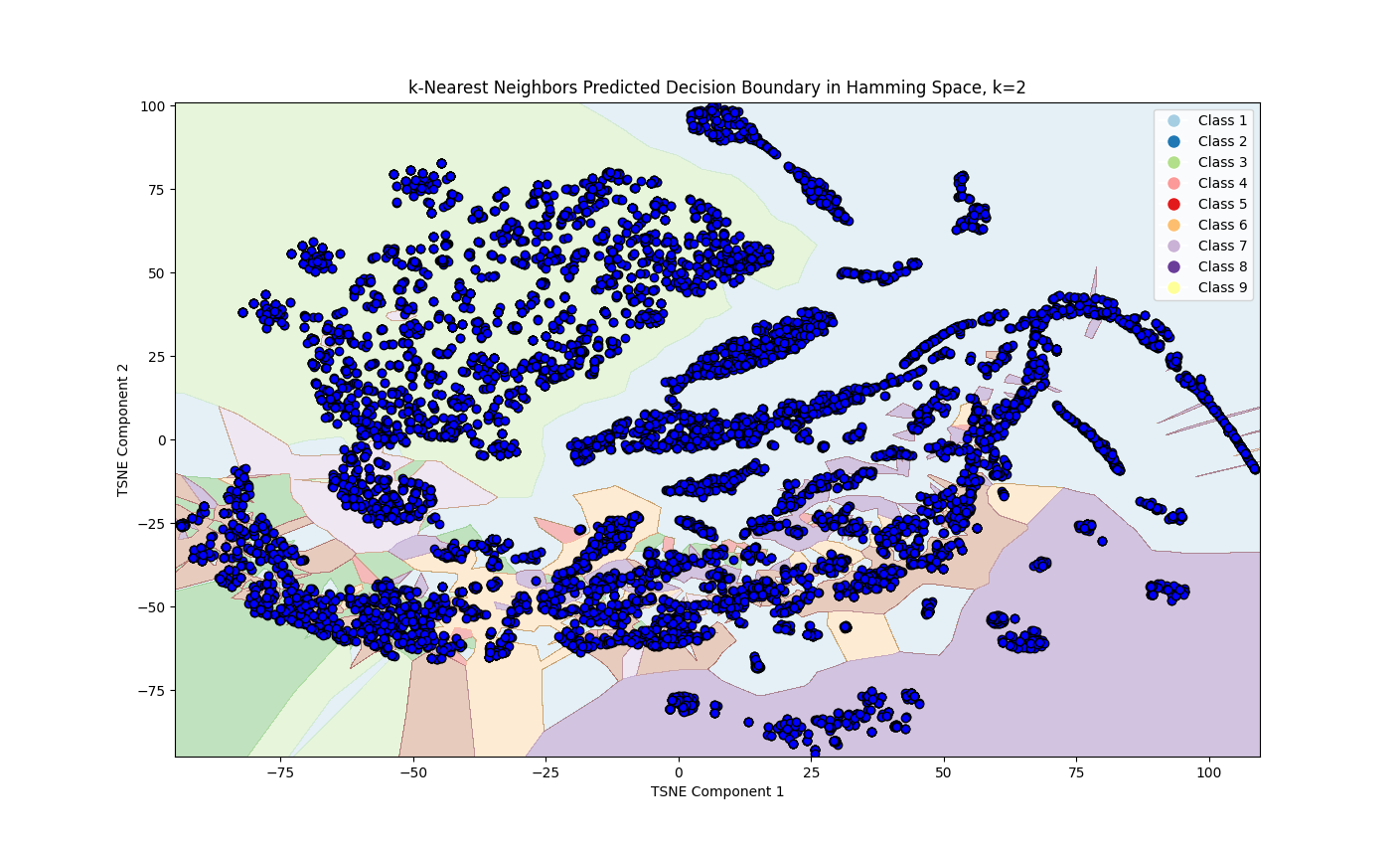}
    \caption{Decision boundary for kNN classifier, k=2.  The decision boundary is plotted in the 2 dimensional projection of the Hamming space using t-SNE.}
    \label{fig:enter-label}
\end{figure*}

\section{Discussion}
The findings of this study show that the kNN classifiers observed have a recorded accuracy that is high overall. For kNN with k=1, the total accuracy measured in 0.875, k=2, has a total accuracy score of 0.871. The impact of k on the total accuracy is shown in figure 7. Values of k less than 4 have an accuracy above 0.850. We can see that as the value of k increases, the total accuracy decreases. Values 5 to 18 have accuracy above 0.800. This decrease slows between k values 10 to 20, which are between the range 0.820 and 0.790.

Likewise, for k=2 the per-class accuracy indicates strong performance, however classes 1 and 5 have the weakest overall metrics as shown in Figure 4 and Table 2. The remaining classes have high precision and recall values based on their predictions across the test set. Accuracy metrics for class 8 are more moderate compared to other classes. Class 1 has lower precision and recall, however these values are balanced. Class 5 has moderate precision, and a low recall value. This indicates that class 5 has a relatively high number of false negatives. Class 3 has the highest values for both precision and recall for this classifier.

Since malicious programs would be encountered in an adversarial environment, the goal of the attacker would be a high rate of false negative classification. The results of classes 1 and 5 having moderate to low recall indicate that this class of malware is misclassified due to the impact of false negatives on classification. A more fine grained analysis of this class of malware can measure the functional overlap of specific samples using the same feature representation and metric space.

A number of trade-offs exist within our method that require further consideration.  One strength is that the similarity of individual samples can be computed directly with no training through the use of the distance metric.  However, in order to evaluate a new sample in comparison to an existing corpus, features must be extracted and collected into a library of examples.  This feature extraction process is expensive in terms of computation time, but can be performed ahead of time by processing the dataset offline.  The feature extraction process described entails segmentation of each binary into basic block segments, then extracting graphs of data dependencies that exist between operands for each segment.  For more detail, see Musgrave et al. 2024. One consideration is that after features have been extracted the metric space must be re-computed for new data.  However, this process has low computational resource requirements, requiring a single matrix multiplication in the case of one-hot encoded vectors.  This is ideal for encountering a finite number of new examples, and can be computed in real-time.  The resulting metric space is high dimensional, but non-parametric methods may be used effectively. Methods sensitive to high dimensions may suffer a decrease in performance. However, support vector machines are able to be used efficiently in high dimensional spaces \cite{musgrave2024search}.

k-Nearest Neighbors as a classification approach has less implementation complexity and relies heavily on the feature representation and distance metric. This was done to test the effectiveness of the  distance metric and feature representation and their ability to capture semantic information. The distance metric is the Hamming distance between two vectors, the vectors are one-hot encoded representations of categorical values. The categorical features are hash representations of graph isomorphic patterns. The graphs were extracted from operand dependencies between data movement instructions, which are hashed for isomorphic uniqueness. The feature representation allows us to use the distance metric to measure similarity. Existing approaches trained on $tf-idf$ features often do not take into account a fine-grained analysis of data dependency graph features, and therefore are required to show that their classification is correlated to operational semantics and structure. Resolution increases in the feature space for $tf-idf$ features present a significant challenge, and there is not a straightforward approach to analyzing the body of the term frequency distribution, which is positively skewed towards data movement instructions. The representation we have chosen addresses these concerns, and provides a foundation for further inferences, since it is constructed in a bottom up approach and tied directly to patterns of data movement \cite{brownlee2017one}, \cite{hamming1950error}.

One explanation for the accuracy measurements in classification is the choice of feature representation, and its ability to capture semantic information. If the data dependency graphs contained no semantic information, then we would expect the accuracy numbers to be very low. Low classification accuracy would be an indication that isomorphism of data dependency graphs between operands of instructions are not good representations of program behavior, operational semantics, or structural patterns. However, an analysis only based on operands of '$mov$' instructions can provide accuracy of 87.5\% with non-parametric learning approaches, which can be increased by analyzing additional instructions. This indicates that the feature representation is sufficiently capturing semantic information.

Accurate classification results for this feature representation provide evidence that direct inferences can be made from the extracted features, since the accuracy corresponds to the ground-truth labels.  Similarity in the metric space corresponds to a direct inference of both the operational semantics and structural properties captured by the feature representation.  This is because data dependency graphs capture both structural and semantic information.  This is validated by the classification accuracy.  Increases in accuracy represent the ability to make inferences that are based on quantifying patterns of operational semantics, behavior, and structural composition.  Based on the presence of patterns of data movement and structural measurements, we are able to make inferences to the patterns of behavior within the program.  This provides the ability for fine-grained analysis of program semantics, and the ability to categorize larger classes of program behavior.

Some limitations in our study are that only operand dependencies between '$mov$' instructions were taken into consideration when constructing the features. This could be expanded to include other instruction types for a more complete picture of data dependencies. This would be an area to provide resolution increases. Additionally, refining the metric space to be more granular will likely provide more insights. Expanding the dataset to include benign samples will likely provide better generalization.

We highlight that our method allows for further refinements, reduction in the search space, and increases in resolution to provide more fine grained results. These approaches are recommended for future studies. In future work we intend to increase the accuracy through increasing the feature resolution, expanding the instructions analyzed, as well as explore the impact of fuzzy systems.

\begin{figure*}
    \centering
    \includegraphics[width=1\textwidth, height=0.6\textwidth]{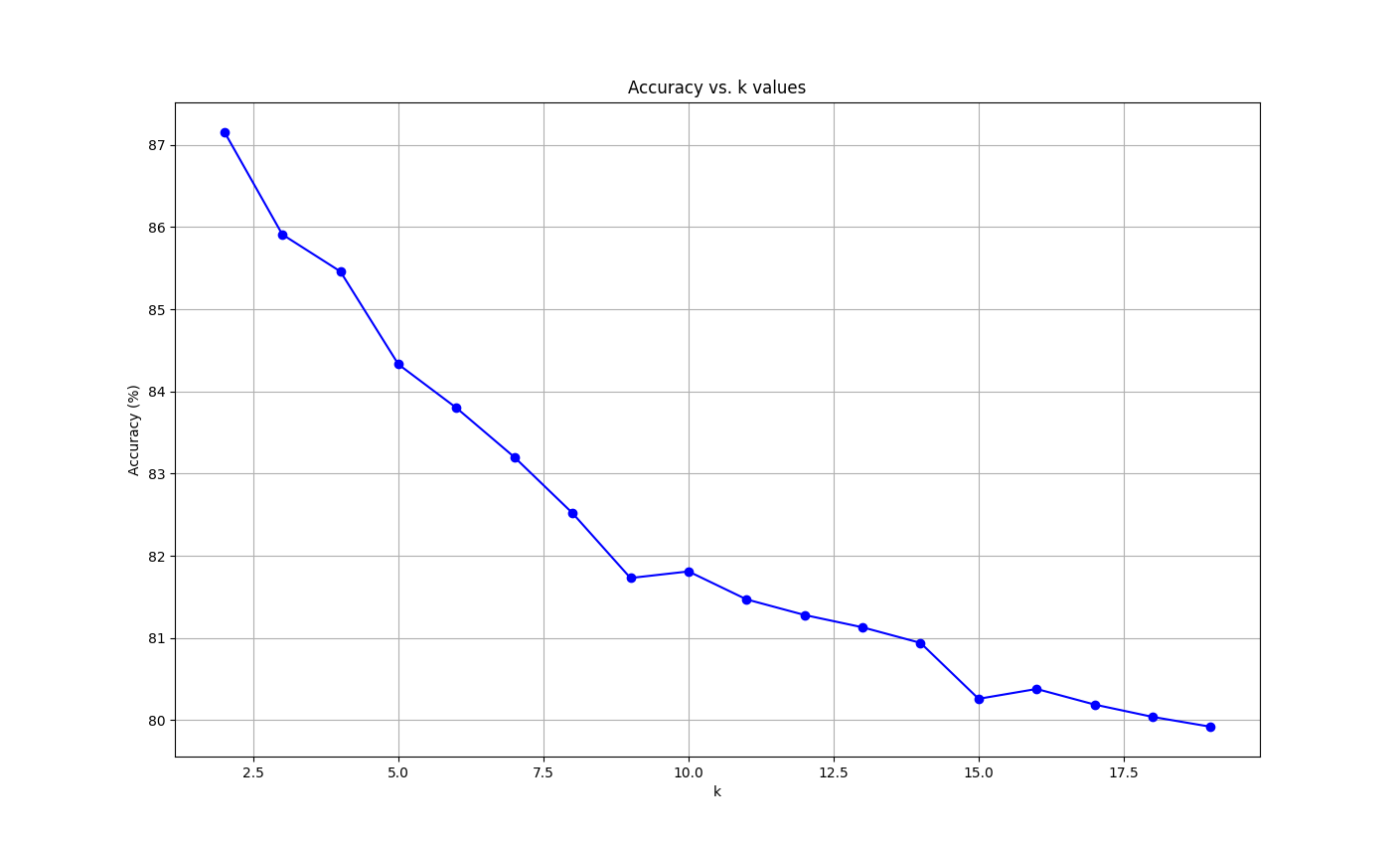}
    \caption{Total accuracy as values of k increase.  Accuracy shown for values of $k$ in the range 2 to 19.}
    \label{fig:enter-label}
\end{figure*}

\section{Conclusion}
In this study we have presented results from k-Nearest Neighbors classifiers across several k values, and have evaluated these classifiers based on precision, recall, specificity, F1 score, and total accuracy. We have shown that a classifier trained on our feature representation for a single instruction obtained a total accuracy for a multi-class classification task of 87.5\%, which can be repeated for additional instructions to gain an increase in accuracy. Similarity in the metric space can be calculated without prior training. This demonstrates the method of feature representation, and validates the hypothesis that data dependency isomorphism is representative of program behavior and operational semantics.

\section*{Acknowledgment}
The authors would like to acknowledge the support of the Air Force Research Lab at Wright-Patterson Air Force Base in Dayton, Ohio.

This study was performed with computing resources provided by the Computer Science department at The College of Wooster in Wooster, Ohio.

\vspace{12pt}

\begin{figure*}
    \centering
    \includegraphics[width=1\textwidth, height=0.65\textwidth]{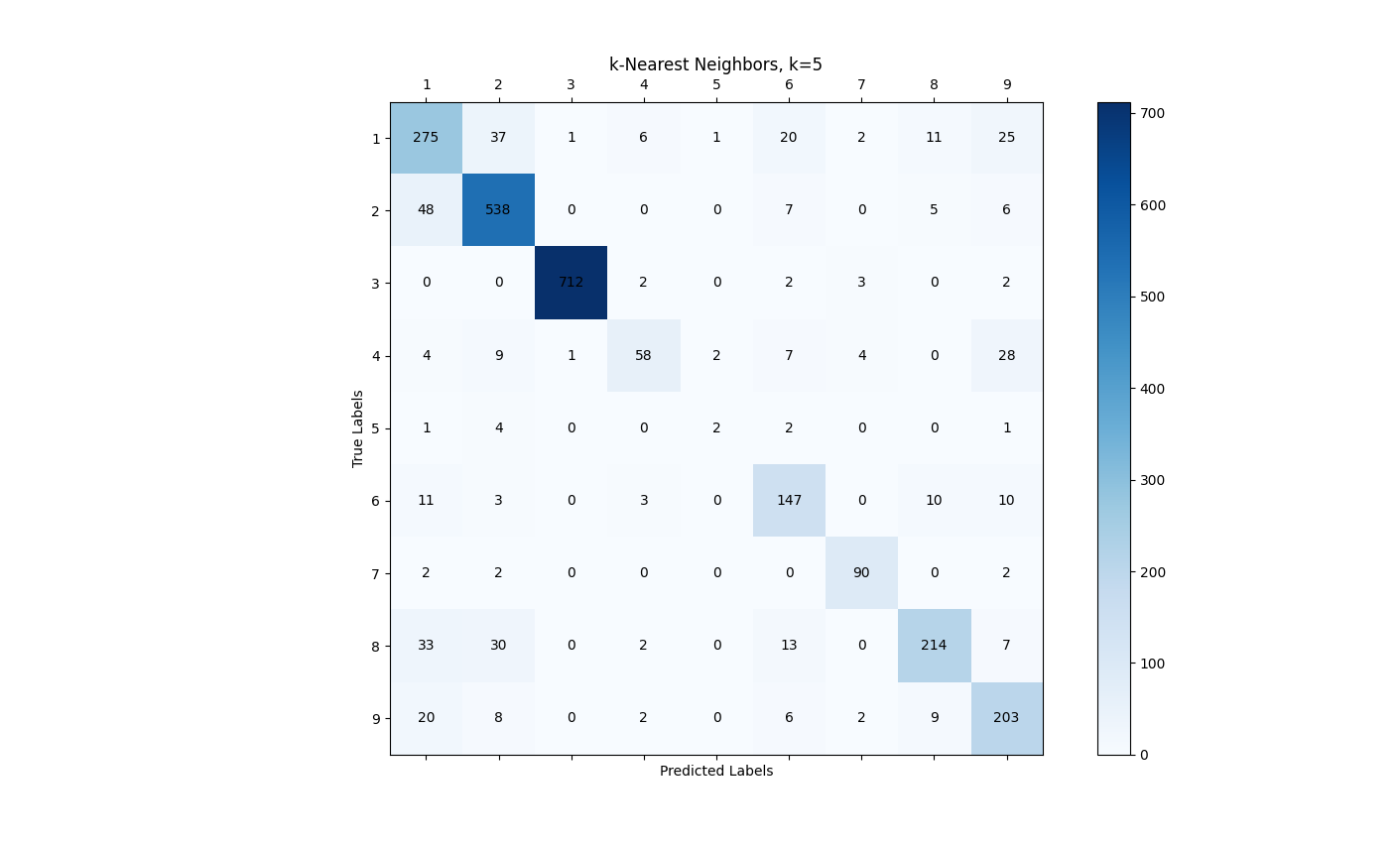}
    \caption{Confusion Matrix for kNN classifier, k=5}
    \label{fig:enter-label}
    \vspace{20pt}
\end{figure*}

\begin{table*}[h]
    \centering
    \begin{tabular}{|c|c|c|c|c|c|}
        \hline
        Class & Accuracy & Precision & Recall & Specificity & F1 Score \\
        \hline
        1 & 0.9306 & 0.6597 & 0.6713 & 0.9868 & 0.6655 \\
        2 & 0.9635 & 0.8074 & 0.8691 & 0.9836 & 0.8371 \\
        3 & 0.9886 & 0.9606 & 0.9916 & 0.9900 & 0.9759 \\
        4 & 0.9739 & 0.8449 & 0.7284 & 0.9935 & 0.7823 \\
        5 & 0.9870 & 0.6900 & 0.2857 & 0.9954 & 0.4024 \\
        6 & 0.9733 & 0.7925 & 0.7737 & 0.9874 & 0.7829 \\
        7 & 0.9887 & 0.9302 & 0.9375 & 0.9930 & 0.9338 \\
        8 & 0.9485 & 0.7347 & 0.7470 & 0.9810 & 0.7408 \\
        9 & 0.9679 & 0.7969 & 0.8133 & 0.9920 & 0.8050 \\
        \hline
    \end{tabular}
    \caption{Accuracy metrics per class for kNN classifier with k=5.  Precision, recall, specificity, and F1 scores are displayed for each class.}
    \label{tab:metrics_k5}
\end{table*}

\begin{figure*}
    \centering
    \includegraphics[width=1\textwidth, height=0.65\textwidth]{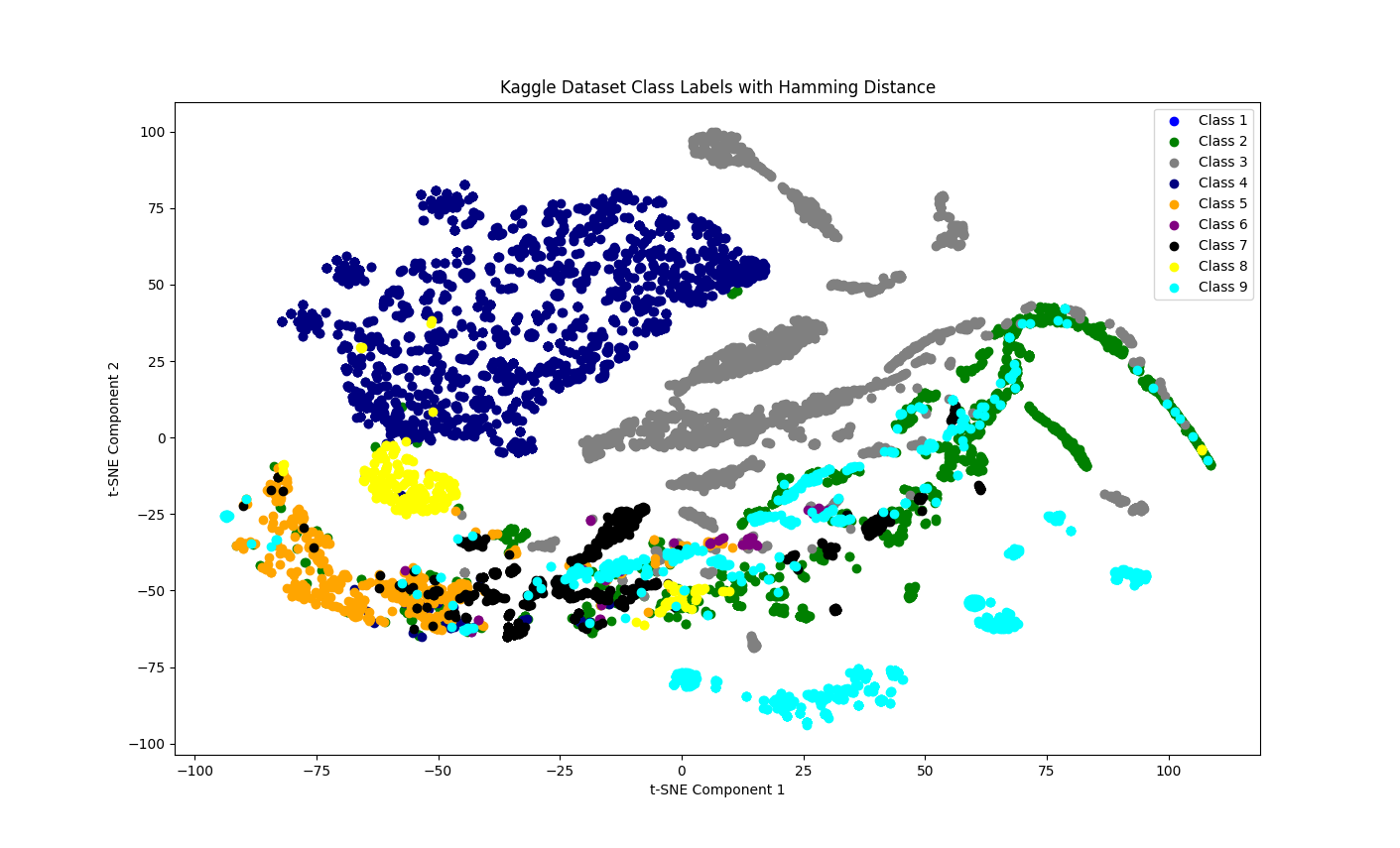}
    \caption{Kaggle Malware Dataset with true class labels.  Picture is the dataset composition in 2 dimensions with true class labels indicated by the color of each data point.}
    \label{fig:enter-label}

    \includegraphics[width=1\textwidth, height=0.65\textwidth]{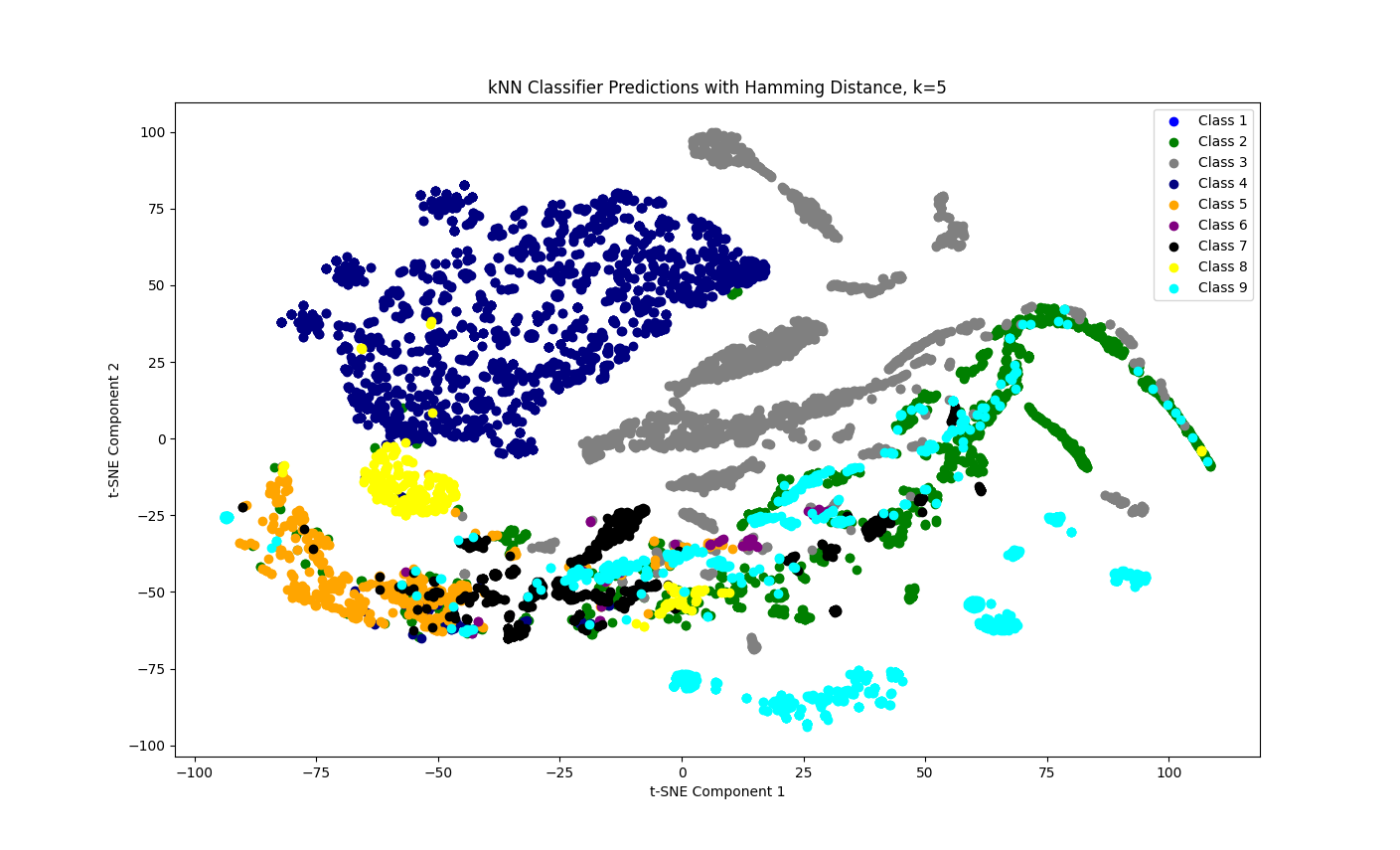}
    \caption{kNN Classifer, k=5}
\end{figure*}


\begin{thebibliography}{00}
\bibitem{souri2018state} A. Souri and R. Hosseini,
\textit{A state-of-the-art survey of malware detection approaches using data mining techniques},
\textit{Human-centric Computing and Information Sciences},
vol. 8, no. 1, pp. 3, 2018.

\bibitem{bruschi2006detecting} D. Bruschi, L. Martignoni, and M. Monga,
\textit{Detecting self-mutating malware using control-flow graph matching},
in \textit{International conference on detection of intrusions and malware, and vulnerability assessment},
pp. 129--143, Springer, 2006.

\bibitem{xu2022systematic} F. F. Xu, U. Alon, G. Neubig, and V. J. Hellendoorn,
  \textit{A systematic evaluation of large language models of code},
  in \textit{Proceedings of the 6th ACM SIGPLAN International Symposium on Machine Programming},
  pp. 1--10, 2022.

\bibitem{kebede2017classification}Kebede, T., Djaneye-Boundjou, O., Narayanan, B., Ralescu, A. \& Kapp, D. Classification of malware programs using autoencoders based deep learning architecture and its application to the microsoft malware classification challenge (big 2015) dataset. {\em 2017 IEEE National Aerospace And Electronics Conference (NAECON)}. pp. 70-75 (2017)

\bibitem{ronen2018microsoft}Ronen, R., Radu, M., Feuerstein, C., Yom-Tov, E. \& Ahmadi, M. Microsoft Malware Classification Challenge. (2018)

\bibitem{cesare2010fast} S. Cesare and Y. Xiang,
\textit{A fast flowgraph based classification system for packed and polymorphic malware on the endhost},
in \textit{2010 24th IEEE International Conference on Advanced Information Networking and Applications},
pp. 721--728, IEEE, 2010.

\bibitem{cesare2013control} S. Cesare, Y. Xiang, and W. Zhou,
\textit{Control flow-based malware variant detection},
\textit{IEEE Transactions on Dependable and Secure Computing},
vol. 11, no. 4, pp. 307--317, 2013.

\bibitem{cesare2010classification} S. Cesare and Y. Xiang,
\textit{Classification of malware using structured control flow},
in \textit{Proceedings of the Eighth Australasian Symposium on Parallel and Distributed Computing-Volume 107},
pp. 61--70, 2010.

\bibitem{shafiq2009pe} M. Z. Shafiq, S. M. Tabish, F. Mirza, and M. Farooq,
\textit{Pe-miner: Mining structural information to detect malicious executables in realtime},
in \textit{International Workshop on Recent Advances in Intrusion Detection},
pp. 121--141, Springer, 2009.

\bibitem{siddiqui2008detecting} M. Siddiqui, M. C. Wang, and J. Lee,
\textit{Detecting trojans using data mining techniques},
in \textit{International Multi Topic Conference},
pp. 400--411, Springer, 2008.

\bibitem{witten1999weka} I. H. Witten, E. Frank, L. E. Trigg, M. A. Hall, G. Holmes, and S. J. Cunningham,
\textit{Weka: Practical machine learning tools and techniques with Java implementations},

\bibitem{chandrasekaran2021malware}Chandrasekaran, M., Ralescu, A., Kapp, D. \& Kebede, T. Malware Detection using the Context of API Calls. {\em NAECON 2021-IEEE National Aerospace And Electronics Conference}. pp. 92-97 (2021)

\bibitem{ledoux2013functracker} C. LeDoux, A. Lakhotia, C. Miles, V. Notani, and A. Pfeffer,
  \textit{$\{$FuncTracker$\}$: Discovering Shared Code to Aid Malware Forensics},
  in \textit{6th USENIX Workshop on Large-Scale Exploits and Emergent Threats (LEET 13)},
  2013.

\bibitem{alrabaee2018fossil} S. Alrabaee, P. Shirani, L. Wang, and M. Debbabi,
\textit{Fossil: a resilient and efficient system for identifying foss functions in malware binaries},
\textit{ACM Transactions on Privacy and Security (TOPS)},
vol. 21, no. 2, pp. 1--34, 2018.
\bibitem{shervashidze2011weisfeiler} N. Shervashidze, P. Schweitzer, E. J. Van Leeuwen, K. Mehlhorn, and K. M. Borgwardt,
\textit{Weisfeiler-lehman graph kernels},
\textit{Journal of Machine Learning Research},
vol. 12, no. 9, 2011.

\bibitem{musgrave2020semantic} J. Musgrave, C. Purdy, A. L. Ralescu, D. Kapp, and T. Kebede,
  \textit{Semantic Feature Discovery of Trojan Malware using Vector Space Kernels},
  in \textit{2020 IEEE 63rd International Midwest Symposium on Circuits and Systems (MWSCAS)},
  pp. 494--499, IEEE, 2020.

\bibitem{musgrave2022latent} J. Musgrave, T. Messay-Kebede, D. Kapp, and A. Ralescu,
  \textit{Latent Semantic Structure in Malicious Programs},
  in \textit{International Conference on Modelling and Development of Intelligent Systems},
  pp. 234--246, Springer, 2022.
  
\bibitem{musgrave2022novel}
  John Musgrave, Temesguen Messay-Kebede, David Kapp, and Anca Ralescu,
  \textit{A Novel Feature Representation for Malware Classification},
  \textit{arXiv preprint arXiv:2210.09580},
  2022.
  
\bibitem{musgrave2024empirical}Musgrave, J., Campan, A., Messay-Kebede, T., Kapp, D. \& Wang, B. Empirical Network Structure of Malicious Programs. {\em Learning}. \textbf{4}, 112 (2024)

\bibitem{hinton2002stochastic} G. E. Hinton and S. Roweis,
  \textit{Stochastic neighbor embedding},
  \textit{Advances in neural information processing systems},
  vol. 15, 2002.


\bibitem{musgrave2024search}Musgrave, J., Campan, A., Messay-Kebede, T., Kapp, D. \& Wang, B. Search and Retrieval in Semantic-Structural Representations of Novel Malware. {\em Advances In Artificial Intelligence And Machine Learning}. \textbf{4}, 117 (2024)


\bibitem{pedregosa2011scikit}Pedregosa, F., Varoquaux, G., Gramfort, A., Michel, V., Thirion, B., Grisel, O., Blondel, M., Prettenhofer, P., Weiss, R., Dubourg, V. \& Others Scikit-learn: Machine learning in Python. {\em The Journal Of Machine Learning Research}. \textbf{12} pp. 2825-2830 (2011)

\bibitem{brownlee2017one} J. Brownlee,
  \textit{Why one-hot encode data in machine learning},
  \textit{Machine Learning Mastery},
  pp. 1--46, 2017.

\bibitem{hamming1950error} R. W. Hamming,
  \textit{Error detecting and error correcting codes},
  \textit{The Bell system technical journal},
  vol. 29, no. 2, pp. 147--160, 1950.

\end{thebibliography}
\end{document}